\documentclass[letter]{aa} 
\usepackage{graphicx}
\usepackage{txfonts}
\usepackage{natbib}
\bibpunct{(}{)}{;}{a}{}{,} 
\def\deg{\ifmmode^\circ\else$^\circ$\fi}
\def\kms{km\thinspace s$^{-1}$}
\def\msun{M$_{\odot}$}

\newcommand{\mum}{\,$\mu$m}

\def\n2h+{N$_2$H$^+$}
\def\c18o{C$^{18}$O}

\def\deg{\ifmmode^\circ\else$^\circ$\fi}
\def\kms{km\thinspace s$^{-1}$}
\def\msun{M$_{\odot}$}

\begin{document}
   \title{The spatial distribution of substellar objects in IC348 and the Orion Trapezium Cluster}

   \subtitle{}

   \author{M. S. N. Kumar
          \and S. Schmeja
		}


   \institute{Centro de Astrof\'{i}sica da Universidade do Porto, 
              Rua das Estrelas, 4150-762 Porto, Portugal
	  \\
	  \email{nanda@astro.up.pt; sschmeja@astro.up.pt }}

   \date{}

\abstract
{}{Some theoretical scenarios suggest the formation of brown dwarfs as
ejected stellar embryos in star-forming clusters. Such a formation
mechanism can result in different spatial distributions of stars and
substellar objects. We aim to investigate the spatial structure of
stellar and substellar objects in two well sampled and nearby embedded
clusters, namely IC348 and the Orion Trapezium Cluster (OTC) to test
this hypothesis.}{Deep near-infrared K-band data complete enough to
sample the substellar population in IC348 and OTC are obtained from
the literature. The spatial distribution of the K-band point sources
is analysed using the Minimum Spanning Tree (MST) method. The Q
parameter and the spanning trees are evaluated for stellar and
substellar objects as a function of cluster core radius R$_c$.}{The
stellar population in both IC348 and OTC display a clustered
distribution whereas the substellar population is distributed
homogeneously in space within twice the cluster core radius. Although
the substellar objects do not appear to be bound by the cluster
potential well, they are still within the limits of the cluster and
not significantly displaced from their birth sites.}{ The spatially
homogeneous distribution of substellar objects is best explained by
assuming higher initial velocities, distributed in a random manner and
going through multiple interactions. The overall spatial coincidence
of these objects with the cluster locations can be understood if these
objects are nevertheless travelling slowly enough so as to feel the
gravitational influence of the cluster. The observations support the
scenario of formation of substellar objects as ``ejected stellar
embryos''. Higher ejection velocities are necessary but net
spatial displacements may not be necessary to explain the
observational data.}


   \maketitle
%

\section{Introduction}

Observations of embedded clusters display structure and mass
segregation generally thought to reflect the structure of molecular
clouds from which they were born and the gravitational potential of
the cluster \citep{lada03}. Massive stars are found to be located at
the centers of the clusters and the radial distribution of cluster
members roughly follow King models or power law variations
\citep{hilhar98}. These findings are generally based on observations
sampling the stellar content of the clusters. With the advent of
sensitive infrared detectors, the substellar content has been
effectively unveiled allowing studies of brown dwarfs and planetary
mass objects. {\em In the rest of this paper we shall refer to Brown
Dwarfs and free-floating planetary mass objects as BDs without
distinction}. The BDs are known to be particularly luminous in their
early stages owing to which hundreds of such objects are detected in
near-infrared surveys of embedded clusters \citep{muench03,lucas05}.

The formation of substellar objects can be explained through at least
five alternative mechanisms \citep{whit07}. One of the important
mechanisms, the so-called ``ejected embryos'' scenario, advocates
formation of BDs as a result of premature ejection of protostellar
embryos from multiple systems. This ``ejection scenario'' attributes
higher velocity dispersion and spatial spread to BDs in comparision to
stellar objects \citep{reipurth01,kroupa03}. Observations of the
radial velocities of BDs and stars in Taurus and Chameleon star
forming regions indicate that there is no significant difference in
the velocity dispersion between stars and brown dwarfs
\citep{luhman07}. Young stars and BDs are also thought to be
homogeneously mixed in the Trapezium cluster and Taurus molecular
cloud. Much of these results are obtained by studying a few objects in
localized regions of the cluster or in a region like Taurus where star
formation occurs in relative isolation. In a deep near-infrared study
of the IC348 cluster, \citet{muench03} analysed the radial
distribution of point sources which indicated a relatively homogeneous
distribution of BDs in space compared to the stellar population that
is bound to the cluster potential.

The structure analysis of embedded clusters has gained importance in
recent years, since numerical simulations of the star formation
process on the scales of  molecular clouds are available
\citep[e.g.][]{bate03}, which can be tested against
observational results. Further, to facilitate the structure analysis
of observed data, quantitative methods of statistical analysis have
also been developed. The nearest neighbour method and the minimum spanning
tree (MST) method \citep{cart04} are two statistical methods
that can be applied on observational data to analyse and quantify the
physical parameters of clusters based on structure analysis. The
structures of some star forming clusters have been analysed
using the MST method \citep{cart04,schmeja06}.

IC348 and the Orion Trapezium cluster (hereafter OTC) are two nearby
well-studied star forming regions for which deep near-infrared data
are available. The OTC in particular has been the target of several
deep surveys that aimed to unveil the sample of BDs down to planetary
mass objects. In both, IC348 and OTC, the BD population is well
sampled. These are also richly populated clusters where a
statistically significant sample of substellar objects are
catalogued. In this letter we will analyse the structure of both
IC348 and OTC to study the variations for the stellar and BD
populations.

\section{Data sets and analysis methods}

Near-infrared point source catalogs for the OTC were obtained from the
\citet{muench02} and \citet{lucas05} studies  via the on-line VizieR
database. The \citet{muench02} data cover a region centered on the
Trapezium cluster to a 10$\sigma$ detection limit of $K=18.1$~mag and
include all the sources found by \citet{hillenbrand00}. The
\citet{lucas05} data obtained through FLAMINGOS on GEMINI south
telescope cover a region adjacent to the boundary of the Trapezium
cluster which suffers from relatively little contamination from the
K-band nebulosity. We combined both the catalogs taking care of
overlaps that resulted in a coverage of $\sim0\fdg1 \times 0\fdg1$.
In the overlapping regions between the two surveys we chose to use the
magnitudes from \citet{muench02}. Data for the IC348 region were
obtained from the FLAMINGOS survey described by \citet{muench03}. The
data cover an $\sim$21\arcmin $\times$21\arcmin\ area centered on the
IC348 cluster to a depth of K$\sim$17~mag.

Based on comparisons to pre-main sequence evolutionary tracks
\citep{lucas05} and luminosity function modelling \citep{muench02},
in the OTC sources brighter than K=14~mag can be considered stars and
sources with K$>$14 can be considered BDs. Similarly, in IC348,
sources fainter than K=15~mag should be complete to 0.04\msun\ brown
dwarfs seen through an extinction of A$_v \sim$7. These estimations
take into account the effects of extinction and nebulosity, hence they
represent a sample not contaminated by such effects. We will use this
classification in the following analysis to distinguish stars and BDs.

  \begin{figure*} \centering 
\includegraphics[width=19cm]{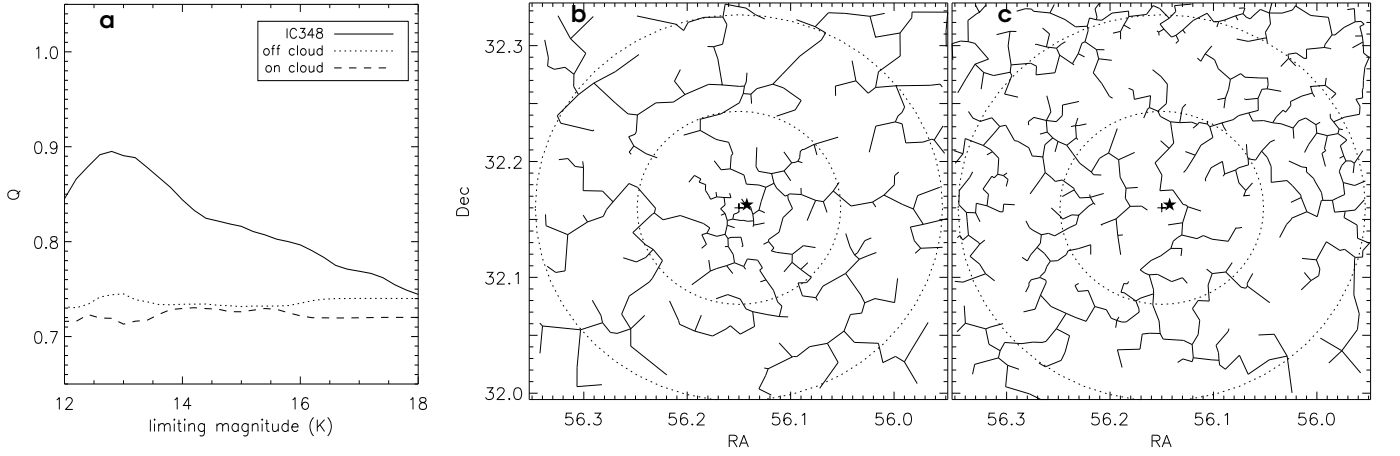}
   
   \caption{IC348: a) Plot showing the variation of the $Q$ parameter with
   magnitude of the point sources considered for the MST analysis. The
   solid line represents the curve for Flamingos K band data complete
   to K=18~mag. The dotted and dashed curves are for two control
   fields data obtained from 2MASS.  b) The minimum spanning trees of
   stars (sources brighter than K$=$15~mag). c) The minimal spanning
   tree for BDs (sources fainter than K$=$15~mag). The inner and outer
   dotted circles have radii of 5\arcmin and 10\arcmin in b) and c).
   }

  \end{figure*}

  \begin{figure*} \centering 
\includegraphics[width=19cm]{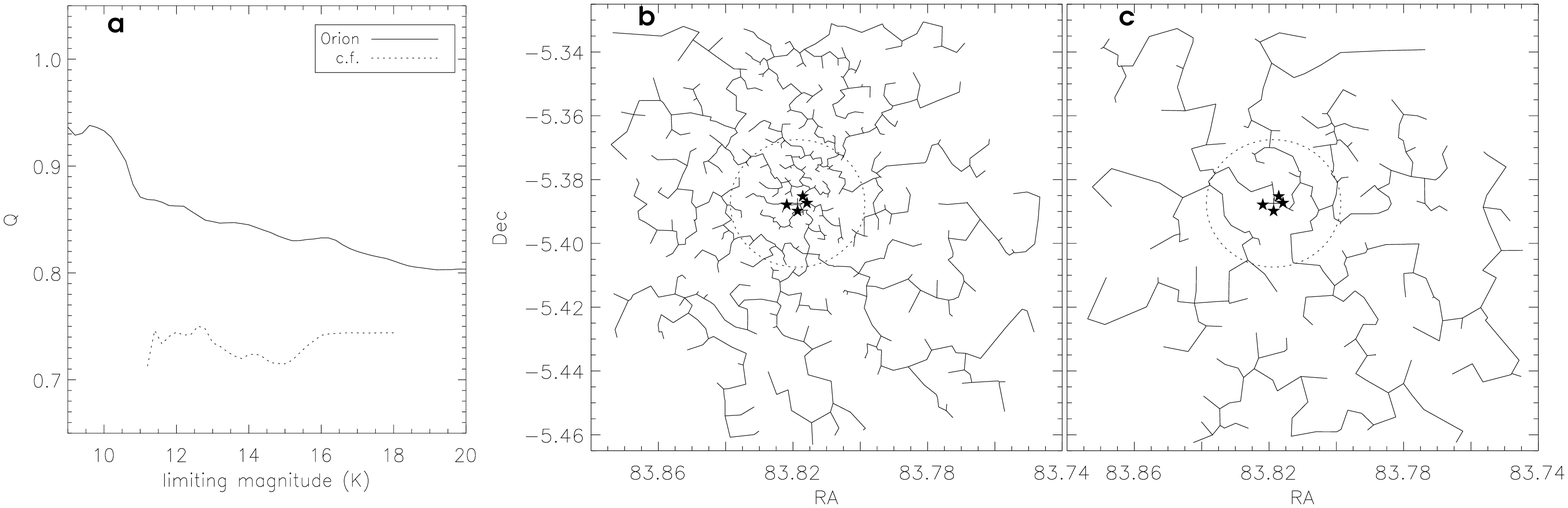}
 
   \caption{Same as Fig.~1 for the Orion Trapezium Cluster. In this
   case the sources brighter than K$=$14~mag are considered as
   stars. The dotted circle marks a radius of 1\farcm2 representing
   the Trapezium cluster radius obtained by a King model fit. } 
   \end{figure*}

Control field data were obtained from the 2MASS point source
catalog. The control fields were selected by looking at large area
(5\deg) IRAS 100\mum\ maps and choosing a region of very low
emission. Point sources within a circle of radius 15\arcmin\, were
extracted with the criteria of photometric quality flags better than
``D'' and without confusion (ccflag=0). The Orion control field is
centered at 05$^h$24$^m$ $-$07$\deg$20$\arcmin$. In the case of IC348,
the off-cloud control field is located at 03$^h$36$^m$
$+$35$\deg$00$\arcmin$; another field on the Perseus molecular
cloud centered at 03$^h$39$^m$ $+$31$\deg$20$\arcmin$ was also
examined. Although the 2MASS survey 100\% completeness limit is at
K=14~mag, several reliable detections of K=15-17~mag are also found in
the point source catalog at a lower $\sigma$ level.

We use the positions and magnitudes to analyse the spatial
distribution of the objects by means of a minimum spanning tree, the
unique set of straight lines (``edges'') connecting a given set of
points without closed loops, such that the sum of the edge lengths is
a minimum \citep[e.g][]{gower69}. We derive the mean edge
length $\ell_{\rm MST}$ as well as the parameter $Q = \bar{\ell}_{\rm
MST}/\bar{s}$ by combining the normalised mean edge length and the
normalised correlation length $\bar{s}$ as described in
\citet{cart04} and \citet{schmeja06}.  The $Q$ parameter quantifies
the cluster structure and distinguishes between a centrally
condensated cluster ($Q > 0.8$) and a cluster showing fractal
substructure ($Q < 0.8$), while $\ell_{\rm MST}$ can be seen as the
mean distance between two neighbouring objects.  We construct the MST
using the algorithm of \citet{prim57} implemented in an IDL routine.

\section{Results}

The MST method was applied on the photometric data to analyse the following:

a) The variation of the $Q$ parameter as a function of source
magnitude (therefore source mass) selected to construct the
tree. At every magnitude (intervals of 0.2~mag), $Q$ was evaluated by
considering all sources brighter than the given magnitude. This
variation was studied for the cluster field and the control fields.

b) The structure of the minimum spanning tree for stars and BDs
separately.  The resultant mean edge lengths $\ell_{\rm MST}$ along
with their standard deviation and minimum and maximum values were
obtained for each tree. Projected separations less than 150\,AU
(indicating source binarity) are neglected in computing $\ell_{\rm
MST}$ to remove their effect on the overall distribution. Further, the
edge lenghts were examined inside and outside a radius that define the
cluster core boundaries.

In Figures~1 and 2 we show these results for IC348 and OTC.  In both
figures, panel a) shows the variation of $Q$ with magnitude, while
panels b) and c) show the tree structure for stellar and substellar
sources respectively. The dotted circles represent boundary limits of
the cluster as defined by radial profile modelling. The $\ell_{\rm
MST}$ values for each zone enclosed by the boundaries, along with the
statistics are listed in Table~1.

It can be seen from Figs.~1a and 2a that the $Q$ values peak at
brighter magnitudes and show a turn-over at K=14~mag for IC348 and
K=11~mag for OTC decreasing smoothly towards fainter magnitudes. In
comparision, the data from the control fields do not show significant
variations of the $Q$ paramater with magnitude. They yield values in
the range of $Q \approx 0.72$, the value expected for a random
distribution. The peak $Q$ values for IC348 and OTC are 0.9 and 0.94,
respectively, indicating centrally condensed configurations. Following
the findings of \citet{cart04} (condensation for $Q>$ 0.8, fractral
substructure for $Q<$ 0.8), these results imply that in IC348 sources
fainter than K=15.5~mag (the point where $Q$ touches 0.8 in Fig.~1a)
are distributed in space with a fractal structure while brigther
sources are more centrally condensed. The same intrepretation in OTC
suggests a general central condensation that is only weak for the
fainter sources.

The structure of the MST in Figs.~1b and 2b display clusters at the
center of the field while Figs.~1c and 2c display a homogeneous
distribution of sources. The clustering can be seen as branches of
smaller edge lengths at the center spreading into branches of longer
edge lengths away from the center. It can also be noted that the
clusters in Figs.~1b and 2b coincide well with the B star in IC348 and
with the Trapezium stars in the OTC cluster, shown with filled star
symbols. The fainter sources do not show detectable variations in the
tree structure across the field. These results are demonstrated more
quantitatively in Table~1. The linear lengths in Table~1 are derived
from the angular lengths by assuming a distance of 315~pc and 500~pc
to IC348 and OTC, respectively. The edge lengths for stars and BDs are
listed individually for each cluster inside and outside the cluster
full-width half maximum boundaries (also referred as cluster core radius
R$_c$). In the IC348 cluster, the boundaries of 5\arcmin and
10\arcmin\ represent the cluster core radius R$_c$ and the limit at
which the radial profile merges with the background. These values are
adopted from \citet{muench02} where the radial distribution of sources
is modelled. Similarly, a boundary of R$_c$=$1\farcm2$ is assumed for
OTC based on the King Model fits from
\citet{hilhar98}.  In IC348, the $\ell_{\rm MST}$ values for stars
within R$_c$=5\arcmin and outside vary by a factor of 1.9,
whereas for substellar objects (denoted as BDs) they show similar
values. In the case of OTC the $\ell_{\rm MST}$ values of stars change
by a factor of 2.12 and those of BDs by 1.4 within and outside the
assumed cluster core radius of R$_c=1\farcm2$.

\begin{table} 
  \caption[]{Minimum Spanning Tree parameters for IC348 and OTC}
  \begin{tabular}{@{}lrrrrr@{}} \hline \hline & $\ell_{\rm MST}$ &
  $\sigma_{\ell}$ & $\ell_{\rm min}$ & $\ell_{\rm max}$ & n$_{\star}$\\ Survey & 10$^3$AU &
  10$^3$AU & 10$^3$ AU & 10$^3$ AU & \\
\hline
IC348 & & & & &\\
\hline
Stars ($<$5\arcmin) & 7.8 & 4.1 & 1.4 & 22.8 & 135\\
Stars ($>$5\arcmin) & 14.7 & 7.7 & 0.8 & 43.8 & 238\\
BDs ($<$5\arcmin) & 9.6 & 5.0 & 1.1 & 25.3 & 98\\
BDs ($>$5\arcmin) & 9.5 & 5.1 & 1.0 & 36.3 & 536\\
\hline
Orion Trapezium & & & & & \\
\hline
Stars ($<$1.2\arcmin) & 2.5 & 1.3 & 0.4 & 6.5 & 228\\
Stars ($>$1.2\arcmin) & 5.4 & 3.4 & 0.16 & 26.1 & 570\\
BDs ($<$1.2\arcmin) & 5.8 & 3.4 & 0.5 & 13.7 & 49\\
BDs ($>$1.2\arcmin) & 8.4 & 5.0 & 0.16  & 26.9 & 258\\
\hline
\hline
\end{tabular}
\end{table}

The above results are free from contamination due to binarity (as we
excluded tree branches smaller than 150\,AU) but include small N
clusters. A total of 19 sources were excluded in the OTC sample that
imitated binary lengths. The comparisions between inside and outside
the cluster radius R$_c$ show that the stars are centrally concentrated in
both IC348 and OTC clusters whereas the BDs are distributed
uniformly. In IC348, the BDs are clearly homogeneously distributed
within the radius of 10\arcmin\ reproducing the results of
\citet{muench03}. The MST  variation of BDs in OTC may
indicate weak concentration at the center, although the region
examined here encompasses only the Trapezium cluster which is the
central part of the much larger Orion Nebula Cluster
\citep{hil97}. Therefore, these variations show, that unlike
stars, the BDs are not limited by the cluster potential in both IC348
and OTC. The substellar objects are rather uniformly distributed
in space imitating a fractal configuration.

\section{Discussion}

The results of the analysis presented in the previous section has two
significant parts:

a) The BDs are spread uniformly in space and not limited by the
cluster potential well whereas stars are bound by a well described
cluster potential. This can be explained if BDs in the cluster are
seen as an N particle system with higher initial velocities than the
stars in random directions  produced by multiple interactions at
early times. The net result will be a relatively uniform distribution
of space positions and velocities instead of being constrained by the
gradient of the cluster potential well. Such a scenario of multiple
objects and chaotic interactions is well described by the numerical
simulations such as that of \citet{bate03}. However, this process
is relevant during the early evolution of the cluster (within
10$^5$ yrs) thus setting up the ``initial'' conditions for BDs. {\em
Higher initial velocities} are necessary to overcome the cluster
potential gradient and produce the homogeneous distribution of spatial
locations. The natural effect of lower mass objects acquiring higher
recoil velocities in comparision to higher mass objects
\citep{reipurth01,kroupa03} can result in BDs having the necessary
ejection velocities. This is in accordance with the predictions of
\citet{kroupa03} who compute higher dispersion in ejection velocities
of 2-3~\kms\ for BDs. 

b) The BDs appear slightly more centrally concentrated in the 1 Myr
old OTC than in the 3 Myr old IC348 cluster. While the BDs show a
homogeneous distribution in space, they are still limited to the
physical boundaries of the cluster or only slightly spread out. This
is evident from the significant number of BDs in comparison to stars
in both clusters (see last column of Table~1). It implies that the BDs
have not traversed in space significantly away from the regions where
they were born.  The more uniform distribution in the 3\,Myr old IC348
cluster with respect to the 1\,Myr old OTC imply that, with time, the
BDs may indeed be moving away from the cluster center, but the net
displacement is not significant.  Assuming higher initial
velocities for BDs, the cluster potential that operates on the larger
scale, will reduce their overall displacement away from the cluster
center. \citet{goodwin05} examine the spatial distribution of BDs
produced by the decay of small N stellar systems as expected from the
embryo ejection scenario and show that significant spatial spread of
BDs with respect to stars may occur only after 5 Myr. Their
predictions are consistent with the results of IC348 and OTC.

The above results are therefore clear signatures of the scenario of
``ejected stellar embryos'' \citep{reipurth01,kroupa03} as an important
mechanism for the formation of brown dwarfs and free-floating
planetary mass objects.

Several studies have been made to investigate the differences in
spatial distribution of stars and BDs in star forming regions. In the
Chamaeleon I cloud, the distribution of BDs and low mass stars was
found to be similar \citep{lopez04} and in the Lupus cloud BDs were
found close to their birth sites \citep{lopez05}.  In the Taurus
molecular cloud, \citet{briceno02} examined the distributions of
nearest neighbours to objects with spectral indices $\le$M6 and $>$M6
and concluded that there were no significant differences in the
spatial distribution of stars and BDs. Examining a smaller group of
T-Tauri stars \citet{martin01} hinted at an anticorrelation between
density of stars and BDs in Taurus. \citet{luhman06} examined an area
of 225 deg$^2$ covering almost the entire Taurus Molecular Cloud and
found no significant differences in the spatial distributions of BDs
and stars.

The above studies were made because according to the ``ejected
embryos'' scenario, BDs may be expected to be significantly
displaced from their birth locations. For example,assuming a 1D
dispersion velocity of 1 \kms they should have moved $\sim$1\,pc away
(corresponding to 7$\arcmin$$\approx$6$\times$R$_c$) from the cluster
in 1Myr old OTC or about 3\,pc away (corresponding to
35$\arcmin$$\approx$7$\times$R$_c$) in the IC348 cluster( if they
do not experience any retarding force from the cluster
potential). However, as evident from the data presented here, the BDs
are indeed located within a region $\sim$ 2$\times$R$_c$ in both the
clusters. Whether there is a significant concentration of BDs much
farther away from the cluster cores in IC348 and OTC is an issue that
needs investigation and is beyond the scope of the data analysed
here. However, following the example of Taurus, where \citet{luhman06}
examined the entire molecular cloud, significant populations of BDs
away from the dense cores may not be expected.

The major difference between previous studies and this work is the
advantage of large number statistics and space densities of point
sources found in IC348 and OTC. This number statistics has made it
possible to effectively apply the MST technique to analyse the spatial
distribution. In the absence of the MST analysis, IC348 and OTC would
imitate the results of Taurus, ChaI or Lupus described above. Indeed
in Fig.~2 of \citet{luhman06} it can be seen that the BDs are sparsely
populated while stars are relatively clustered. MST analysis for
Taurus may also show the kind of variations found here, but the IC348
and OTC are clearly better data sets for effective application of the
MST method which is ``statistical'' in nature. In conclusion, we have
shown that the near-infrared observations of the IC348 and OTC
embedded clusters are in accordance with the ``ejected stellar
embryos'' scenario for the formation of substellar objects. Higher
ejection velocities for substellar objects are essential but their
spatial displacements from the birth sites may not be necessary to
explain the observed data.

\begin{acknowledgements}

 We thank an anonymous referee for useful suggestions. MSNK and SS are
 supported by a research grant POCTI/CFE-AST/55691/2004 approved by
 FCT and POCTI, with funds from the European community programme
 FEDER. This research has made use of the SIMBAD and VizieR databases,
 operated at CDS, Strasbourg, France.  This publication makes use of
 data products from the Two Micron All Sky Survey, which is a joint
 project of the University of Massachusetts and the Infrared
 Processing and Analysis Center/California Institute of Technology,
 funded by the National Aeronautics and Space Administration and the
 National Science Foundation.

\end{acknowledgements}

\end{document}